\PassOptionsToPackage{table}{xcolor}

\documentclass[letterpaper, 10 pt, conference]{class/ieeeconf} 

\usepackage{flushend}
\usepackage{import}
\usepackage{class/tudor}
\usepackage{class/abbrv}

\usepackage{caption}
\usepackage{color}
\usepackage[table,xcdraw]{xcolor}

\definecolor{cite_color}{RGB}{240,150,12}

\makeatletter
\let\NAT@parse\undefined
\makeatother
\usepackage[
bookmarks=false, linkcolor=blue, urlcolor=blue, citecolor=blue
]{hyperref} 

\hypersetup{
    colorlinks=true,
    linkcolor=red,
    filecolor=magenta,      
    urlcolor=blue,
    pdfstartview={FitH},
    citecolor =blue
    }

\IEEEoverridecommandlockouts                              
\title{\LARGE \bf SplineFormer: An Explainable Transformer-Based Approach for Autonomous Endovascular Navigation}

\author{Tudor Jianu$^1$, Shayan Doust$^1$, Mengyun Li$^1$, Baoru Huang$^1$, Tuong Do$^{1}$, Hoan Nguyen$^2$, Karl Bates$^3$, \\ Tung D. Ta$^4$, Sebastiano Fichera$^5$, Pierre Berthet-Rayne$^{6,7}$, Anh Nguyen$^1$
\thanks{$^1$ Department of Computer Science, University of Liverpool, UK {\tt t.jianu@liverpool.ac.uk}}
\thanks{$^2$ University of Information Technology, Vietnam}
\thanks{$^3$ Faculty of Health and Life Sciences, University of Liverpool, UK}
\thanks{$^4$ University of Tokyo, Japan}
\thanks{$^5$ Department of Mechanical, Materials and Aerospace Engineering, University of Liverpool, UK}
\thanks{$^6$ Honorary Fellow, University of Liverpool, UK}
\thanks{$^7$ 3IA Cote d'Azur, Sophia Antipolis, France}}

\begin{document}

\newtheorem{problem}{Problem}
\newtheorem{lemma}{Lemma}
\newtheorem{theorem}[lemma]{Theorem}
\newtheorem{claim}{Claim}
\newtheorem{corollary}[lemma]{Corollary}
\newtheorem{definition}[lemma]{Definition}
\newtheorem{proposition}[lemma]{Proposition}
\newtheorem{remark}[lemma]{Remark}
\newenvironment{LabeledProof}[1]{\noindent{\it Proof of #1: }}{\qed}

\def\beq#1\eeq{\begin{equation}#1\end{equation}}
\def\bea#1\eea{\begin{align}#1\end{align}}
\def\beg#1\eeg{\begin{gather}#1\end{gather}}
\def\beqs#1\eeqs{\begin{equation*}#1\end{equation*}}
\def\beas#1\eeas{\begin{align*}#1\end{align*}}
\def\begs#1\eegs{\begin{gather*}#1\end{gather*}}

\newcommand{\poly}{\mathrm{poly}}
\newcommand{\eps}{\epsilon}
\newcommand{\e}{\epsilon}
\newcommand{\polylog}{\mathrm{polylog}}
\newcommand{\rob}[1]{\left( #1 \right)} 
\newcommand{\sqb}[1]{\left[ #1 \right]} 
\newcommand{\cub}[1]{\left\{ #1 \right\} } 
\newcommand{\rb}[1]{\left( #1 \right)} 
\newcommand{\abs}[1]{\left| #1 \right|} 
\newcommand{\zo}{\{0, 1\}}
\newcommand{\zonzo}{\zo^n \to \zo}
\newcommand{\zokzo}{\zo^k \to \zo}
\newcommand{\zot}{\{0,1,2\}}
\newcommand{\en}[1]{\marginpar{\textbf{#1}}}
\newcommand{\efn}[1]{\footnote{\textbf{#1}}}
\newcommand{\vecbm}[1]{\boldmath{#1}} 
\newcommand{\uvec}[1]{\hat{\vec{#1}}}
\newcommand{\thv}{\vecbm{\theta}}
\newcommand{\junk}[1]{}
\newcommand{\var}{\mathop{\mathrm{var}}}
\newcommand{\rank}{\mathop{\mathrm{rank}}}
\newcommand{\diag}{\mathop{\mathrm{diag}}}
\newcommand{\tr}{\mathop{\mathrm{tr}}}
\newcommand{\acos}{\mathop{\mathrm{acos}}}
\newcommand{\atantwo}{\mathop{\mathrm{atan2}}}
\newcommand{\SVD}{\mathop{\mathrm{SVD}}}
\newcommand{\quadf}{\mathop{\mathrm{q}}}
\newcommand{\linterp}{\mathop{\mathrm{l}}}
\newcommand{\sgn}{\mathop{\mathrm{sign}}}
\newcommand{\sym}{\mathop{\mathrm{sym}}}
\newcommand{\avg}{\mathop{\mathrm{avg}}}
\newcommand{\mean}{\mathop{\mathrm{mean}}}
\newcommand{\erf}{\mathop{\mathrm{erf}}}
\newcommand{\grad}{\nabla}
\newcommand{\R}{\mathbb{R}}
\newcommand{\defeq}{\triangleq}
\newcommand{\dims}[2]{[#1\!\times\!#2]}
\newcommand{\sdims}[2]{\mathsmaller{#1\!\times\!#2}}
\newcommand{\udims}[3]{#1}
\newcommand{\udimst}[4]{#1}
\newcommand{\com}[1]{\rhd\text{\emph{#1}}}
\newcommand{\ind}{\hspace{1em}}
\newcommand{\argmin}[1]{\underset{#1}{\operatorname{argmin}}}
\newcommand{\floor}[1]{\left\lfloor{#1}\right\rfloor}
\newcommand{\step}[1]{\vspace{0.5em}\noindent{#1}}
\newcommand{\quat}[1]{\ensuremath{\mathring{\mathbf{#1}}}}
\newcommand{\norm}[1]{\left\lVert#1\right\rVert}
\newcommand{\ignore}[1]{}
\newcommand{\specialcell}[2][c]{\begin{tabular}[#1]{@{}c@{}}#2\end{tabular}}
\newcommand*\Let[2]{\State #1 $\gets$ #2}
\newcommand{\algorithmicbreak}{\textbf{break}}
\newcommand{\Break}{\State \algorithmicbreak}
\newcommand{\ra}[1]{\renewcommand{\arraystretch}{#1}}

\renewcommand{\vec}[1]{\mathbf{#1}} 

\algdef{S}[FOR]{ForEach}[1]{\algorithmicforeach\ #1\ \algorithmicdo}
\algnewcommand\algorithmicforeach{\textbf{for each}}
\algrenewcommand\algorithmicrequire{\textbf{Require:}}
\algrenewcommand\algorithmicensure{\textbf{Ensure:}}
\algnewcommand\algorithmicinput{\textbf{Input:}}
\algnewcommand\INPUT{\item[\algorithmicinput]}
\algnewcommand\algorithmicoutput{\textbf{Output:}}
\algnewcommand\OUTPUT{\item[\algorithmicoutput]}

\maketitle
\thispagestyle{empty}
\pagestyle{empty}

\begin{abstract}
Endovascular navigation is a crucial aspect of minimally invasive procedures, where precise control of curvilinear instruments like guidewires is critical for successful interventions. A key challenge in this task is accurately predicting the evolving shape of the guidewire as it navigates through the vasculature, which presents complex deformations due to interactions with the vessel walls. Traditional segmentation methods often fail to provide accurate real-time shape predictions, limiting their effectiveness in highly dynamic environments. To address this, we propose SplineFormer, a new transformer-based architecture, designed specifically to predict the continuous, smooth shape of the guidewire in an explainable way. By leveraging the transformer’s ability, our network effectively captures the intricate bending and twisting of the guidewire, representing it as a spline for greater accuracy and smoothness. We integrate our SplineFormer into an end-to-end robot navigation system by leveraging the condensed information. The experimental results demonstrate that our SplineFormer is able to perform endovascular navigation autonomously and achieves a 50\% success rate when cannulating the brachiocephalic artery on the real robot.


\end{abstract}


\section{Introduction}\label{Sec:Intro}

Cardiovascular diseases remain the leading cause of mortality worldwide, accounting for over a million deaths annually, with coronary heart disease and cerebrovascular disease being the primary contributors~\cite{townsend2016cardiovascular}. Endovascular interventions, including percutaneous coronary intervention (PCI), pulmonary vein isolation (PVI), and mechanical thrombectomy (MT), have become well-established procedures for treating cardiovascular conditions~\cite{thukkani2015endovascular,goyal2016endovascular,giacoppo2017percutaneous,lindgren2018endovascular}. These minimally invasive techniques involve navigating a \textit{guidewire} and \textit{catheter} from an insertion point to the target site within the vasculature, guided by intraoperative fluoroscopy. Upon reaching the target, specific treatments such as thrombus removal, stent deployment, or tissue ablation are performed~\cite{brilakis2020manual}. Despite their effectiveness, the success of endovascular interventions is highly time-sensitive, particularly in acute cases like stroke, where delays beyond \(7.3\) hours can significantly diminish the benefits of MT~\cite{saver2016time}. However, only a small fraction of eligible patients receive these life-saving interventions in time, underscoring the critical need for advancements in procedural efficiency and precision~\cite{mcmeekin2017estimating}.

To achieve the precise navigation required for endovascular interventions, operators rely heavily on fluoroscopic imaging to visualize the vasculature and guide the instruments. However, this reliance on fluoroscopy poses several risks, including radiation exposure to both patients and surgeons, and the potential for nephrotoxicity from contrast agents used during angiography~\cite{rudnick1995nephrotoxicity}. Moreover, the manual operation of catheters and guidewires demands significant skill and dexterity from the operators, which can lead to complications such as vessel perforation and distal embolization~\cite{hausegger2001complications}. These challenges highlight the necessity for improved endovascular robots and automated techniques to enhance the safety and efficacy of endovascular navigation.

\begin{figure}[t]
    \centering
    \includegraphics[width=\linewidth, height=0.45\linewidth]{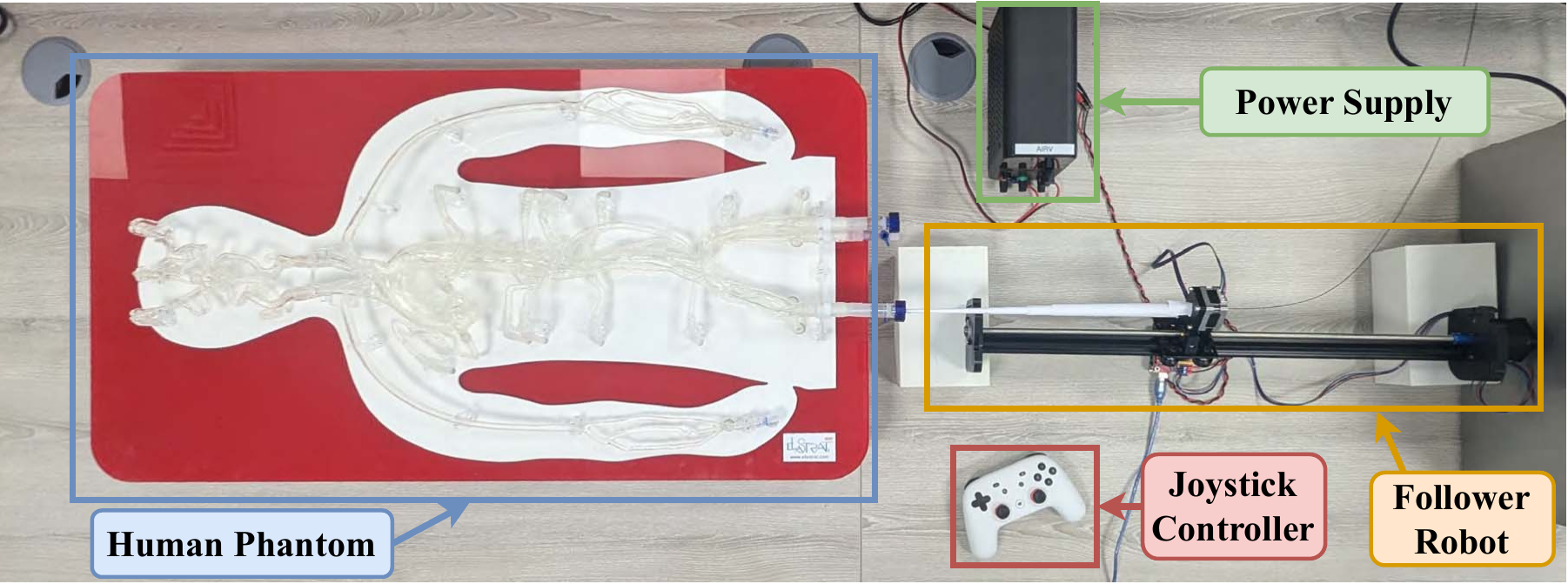}
    \vspace{1mm}
    \caption{\textbf{System Overview:} The experimental setup includes three main components: \textit{i)} an anatomically accurate half-body vascular phantom model from Elastrat Sarl Ltd.; \textit{ii)} a joystick controller (robotic leader); and \textit{iii)} a robotic follower. Data is collected through teleoperation, where the robotic follower is controlled by the joystick controller.
    }
    \label{fig:intro}
\end{figure}

Recent advancements in machine learning offer promising results for addressing the challenges in endovascular interventions~\cite{berthet2023isnake}. Among various learning techniques, Reinforcement Learning (RL) has emerged as a promising method for enhancing the precision and efficiency of these procedures. Various RL algorithms have been explored in endovascular navigation, demonstrating the ability to learn complex control strategies~\cite{chi2018trajectory,behr2019deep,kweon2021deep,cho2022sim,karstensen2022learning}. Despite the potential, the application of RL in endovascular interventions is still in its early stages, with most studies were conducted in controlled environments. The high cognitive workload on operators, the need for specialized skills to interact with robotic interfaces, and the lack of haptic feedback are notable limitations of existing systems~\cite{mofatteh2021neurosurgery,crinnion2022robotics}. Furthermore, ensuring safety, enhancing explainability, and addressing ethical considerations are paramount as these technologies advance towards fully autonomy~\cite{FOSCHVILLARONGA2021105528}.

In this paper, we introduce \textbf{SplineFormer}, a B-spline transformer model which leverages the guidewire's inherent geometrical structure in an efficient manner. Unlike most current methods that rely on the segmentation mask for navigation~\cite{abdelaziz2019toward,chi2018learning,chi2020collaborative}, our key contribution is an \textit{explainable representation} that can be applied to robots for endovascular interventions. We trial our method in a physical environment via the use of a high-fidelity human phantom and robotic setup (Fig.~\ref{fig:intro}). Utilizing our method, the robot is able to perform endovascular navigation task in a \textit{fully autonomous} manner. Our contributions are summarized as follows:
\begin{itemize} 
    \item
    We present SplineFormer, an explainable transformer-based network capable of inferring the guidewire's geometry in a constrained parametric space.     
    \item
    We train our network to \textit{i)} retrieve meaningful and concise representations, and \textit{ii)} use this latent space to derive the appropriate actions to successfully navigate the anatomy.
\end{itemize}

\section{Related Works}\label{sec:related-works}

\textbf{Guidewire and Catheter Segmentation.} In medical imaging, guidewire and catheter segmentation represents a critical research area due to its relevance in interventional procedures. Traditional methods for segmenting guidewires and catheters often rely on image processing techniques such as pixel intensity, texture analysis, and histogram-based methods~\cite{brost2009_tracking,brost2010respiratory,baert2003_guidewire_tracking,wu2011learning,nguyen2020end}. Techniques like the Hough transform and multi-thresholding primarily focus on detecting local image features~\cite{sheng2009automatic,kao2015_chest_CTT}. Semi-automated approaches, including user-guided input, and curvature-based methods have also been explored to enhance segmentation~\cite{keller2007semi,bismuth2012curvilinear}. More advanced techniques employ multiscale vessel enhancement and adaptive binarization to improve real-time performance~\cite{ma2018novel,kongtongvattana2023shape}. However, these methods are often sensitive to variations in image quality and fail to adequately capture the continuous, smooth geometry of the catheter and guidewire, resulting in inconsistent and fragmented segmentation.

\begin{figure}[t]
     \centering
     \includegraphics[width=0.48\linewidth]{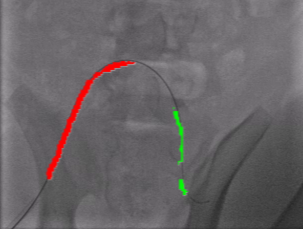}\hfill
     \includegraphics[width=0.48\linewidth]{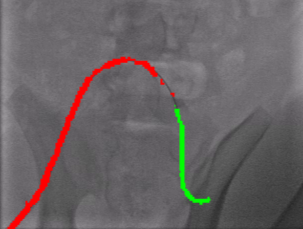}
     \vspace{2ex}
     \caption{\textbf{Segmentation Failure Cases.} Due to challenges in capturing thin and elongated guidewire, segmentation models often produce discontinuous segmented maps.}
     \label{fig:segment-fail}
\end{figure}

Recent advancements in deep learning, have introduced more robust approaches for guidewire segmentation. Convolutional Neural Networks (CNNs) have shown notable improvements by automatically extracting features from medical images~\cite{wang2009robust,chen2016guidewire,wang2017_detection,pauly2010machine,yang2019improving,zaffino2019fully}. Research in this area spans from segmenting the tip, to a dual segmentation of guidewires and catheters~\cite{yang2019improving}. Architectures like U-Net~\cite{ronneberger2015u} have achieved state-of-the-art results in segmentation tasks, and further refinements such as adaptive U-Net and scale-recurrent networks have improved performance in X-ray image sequences~\cite{yi2019_rnn_synthetic,ambrosini2017fully}. Despite these advancements, deep learning models face challenges often struggle to represent the continuous geometry of the guidewires, leading to suboptimal clinical results. The most common form of failure cases is represented by surgical tool fragmentation, as highlighted in numerous studies~\cite{zhang2022real,huang2022self,huang2022simultaneous} and which can be visualized in Fig.~\ref{fig:segment-fail}. This highlights the need for methods capable of preserving the inherent smoothness of the tools while maintaining segmentation accuracy across varying conditions.

\textbf{Autonomous Navigation.} The navigation of surgical tools within the vasculature represents a strenuous task that has seen rapid advancement due to its promising benefits (\ie reduced operative times and diminished radiation exposure). In this area of research, works tend to rely on reinforcement learning to achieve a high level of autonomy~\cite{jianu2024cathsim}. Various RL algorithms have been applied in a significant amount of studies exploring autonomous endovascular navigation~\cite{chi2018learning, chi2020collaborative, behr2019deep, you2019automatic, kweon2021deep, cho2022sim, karstensen2022learning}. Moreover, a significant proportion of studies employed demonstrator data, such as GAIL~\cite{chi2020collaborative}, Behavior Cloning~\cite{chi2018learning,jianu2024autonomous}, or Human Demonstration, as part of Learning from Demonstration (LfD) to augment RL training~\cite{chi2018learning,behr2019deep,kweon2021deep,cho2022sim}. By contrast, the remaining studies relied purely on RL without demonstrator input~\cite{you2019automatic, karstensen2022learning}. While these approaches have demonstrated significant potential, the integration of segmentation as an intermediate step could vastly improve performance, as evidenced by similar advancements in autonomous driving~\cite{zhang2022real}.

\section{Endovascular Robot Setup}\label{sec:robotic-setup}

\begin{figure*}[t]
    \centering
    \subfloat[]{\includegraphics[width=0.37\linewidth]{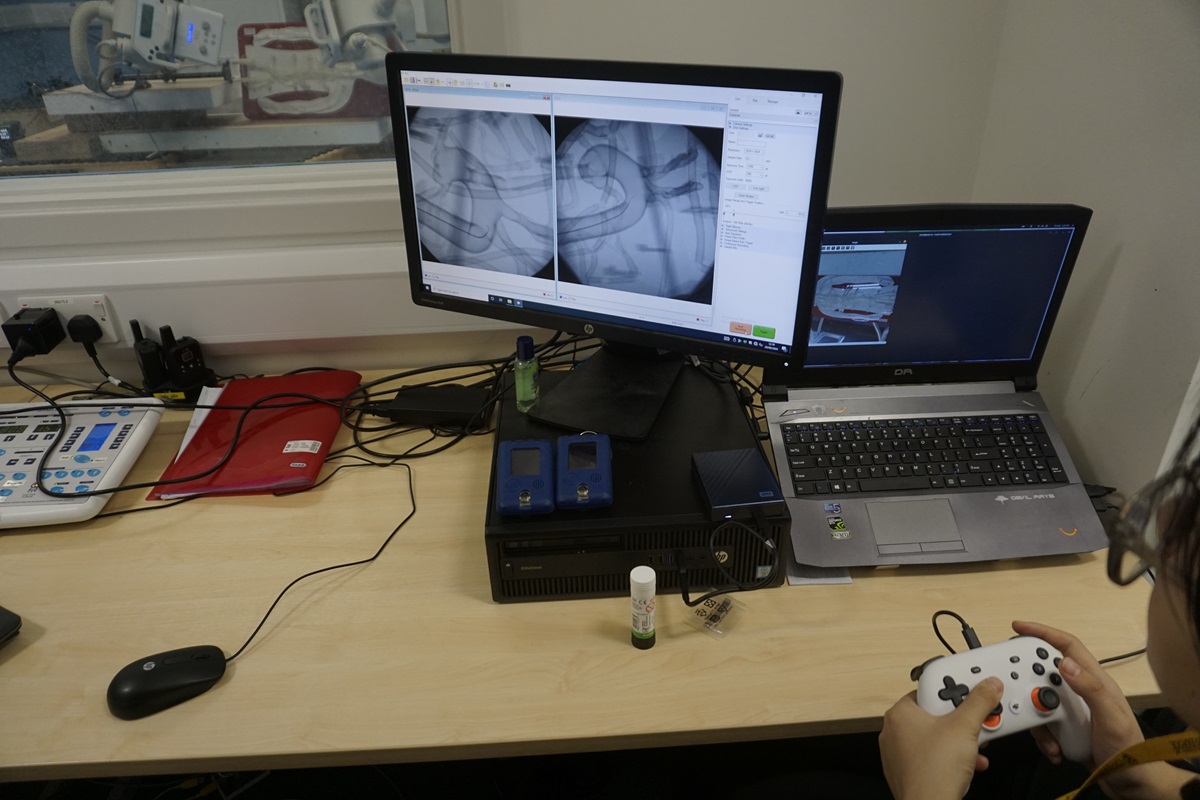}} \hspace{1ex}
    \subfloat[]{\includegraphics[width=0.37\linewidth]{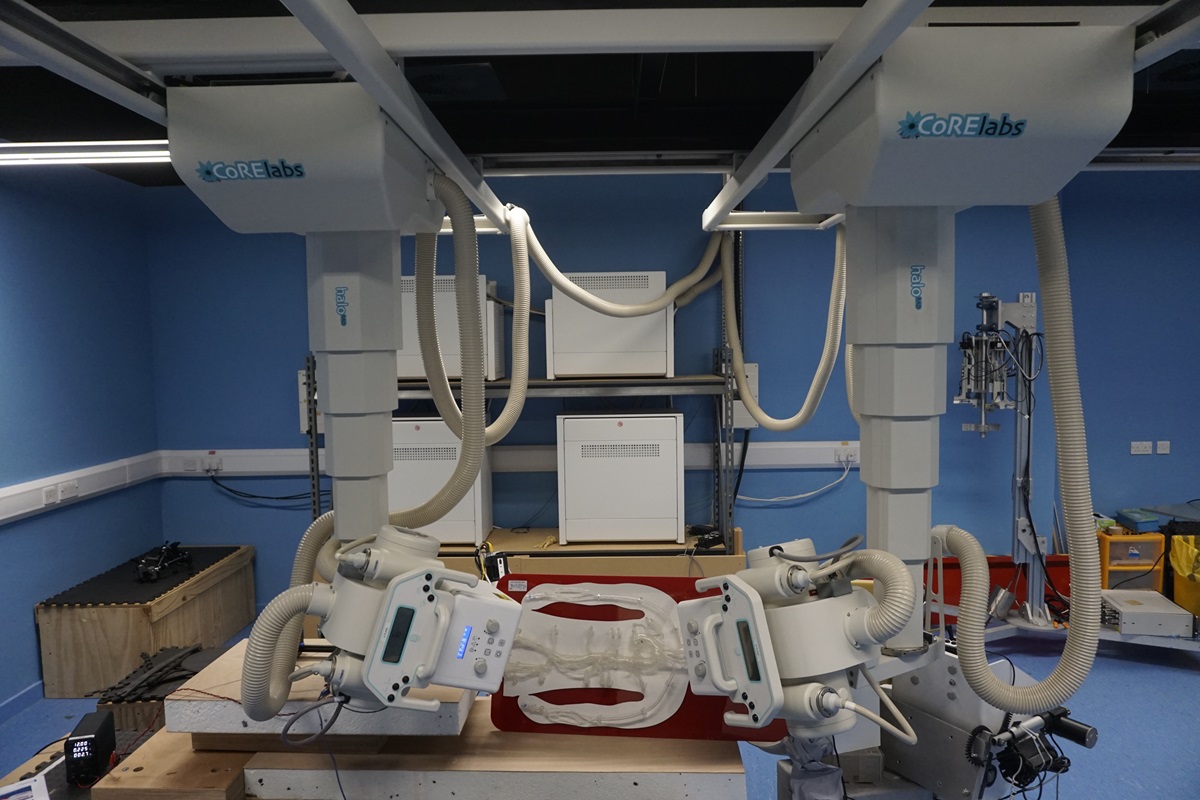}} \hspace{1ex}
    \subfloat[]{\includegraphics[width=0.2\linewidth, height=0.247\linewidth]{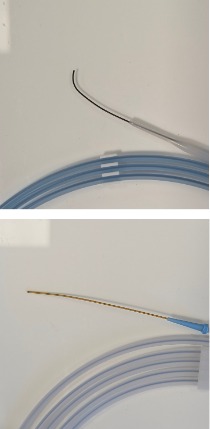}}
    \vspace{2ex}
    \caption{\textbf{Robot Setup in X-ray Room:} The data acquisition and teleoperation process is highlighted, showcasing the user navigating the guidewire towards the designated target within the vascular phantom. The setup employs a Bi-planar X-ray. To enhance data variability, two different guidewires are used—the Radifocus™ Guide Wire M Stiff Type with an angled tip and the Nitrex Guidewire straight tip.}
    \label{fig:materials}
\end{figure*}

\textbf{Robot Setup.} 
Endovascular robotic systems commonly employ a leader-follower (master-slave) architecture, where the master device (controlled by the operator) maps input commands to the slave robot that manipulates the catheter~\cite{Song2023-ns}. These systems often provide up to 6 degrees of freedom (DoF), allowing precise translation and rotation of the catheter~\cite{saliba2006novel}. Human-machine interfaces, such as multi-DoF joysticks or handheld devices, translate operator movements into electromechanical actions, enabling accurate catheter navigation during procedures~\cite{saliba2006novel,khan2013first}. In our system, since we only use a guidewire, the robotic setup focuses solely on translation and rotation movements of the guidewire. This simplifies the design, making it easier to replicate compared to more complex multi-DoF systems. The system consists of a Nema 17 Bipolar \(\SI{59}{\newton\centi\meter}\) \(\SI{2}{\ampere}\) stepper motor for actuation, alongside an additional motor with for rotation. Control is handled by an Arduino Uno Rev3 with a CNC shield and two A4899 drivers, powered by a \(\SI{12}{\volt}\) DC power source, with teleoperation input provided via a Google Stadia joystick.

\textbf{Data Collection.} Our experiments utilize a Bi-planar X-ray system (Fig.~\ref{fig:materials}) equipped with \(\SI{60}{\kilo\watt}\) Epsilon X-ray Generators (EMD Technologies Ltd.) and \(16\)-inch Image Intensifier Tubes (Thales), incorporating dual focal spot Varian X-ray tubes for high-definition imaging. System calibration is achieved using acrylic mirrors and geometric alignment grids. To simulate human vascular anatomy, we employ a half-body vascular phantom model (Elastrat Sarl Ltd., Switzerland) enclosed in a transparent box and integrated into a closed water circuit to replicate blood flow. The model, constructed from soft silicone and featuring continuous flow pumps, was derived from detailed postmortem vascular casts, ensuring anatomical accuracy consistent with human vasculature, as documented in previous studies~\cite{martin1998vitro,gailloud1999vitro}. Finally, we utilized a Radifocus™ Guide Wire M Stiff Type (Terumo Ltd.), a \(\SI{0.89}{\milli\meter}\) nitinol wire with a \(\SI{3}{\centi\meter}\) angled tip and Nitrex Guidewire straight tip.

\textbf{Dataset.} We compiled a dataset of 8,746 high-resolution samples (\(1,024 \times 1,024\) pixels). This dataset includes \(4,373\) paired instances with and without a simulated blood flow medium. Specifically, it comprises \(6,136\) samples from the Radifocus Guidewire and \(2,610\) from the Nitrex guidewire, establishing a robust foundation for automated guidewire tracking in bi-planar scanner images. Manual annotation was performed using the Computer Vision Annotation Tool (CVAT)~\cite{cvat2023}, where polylines were created to accurately track the dynamic path of the guidewire. The dataset is partitioned using a stratified sample method.

\section{Autonomous Endovascular Navigation}~\label{sec:method}
\begin{figure*}[t]
    \centering
    \includegraphics[width=1\linewidth]{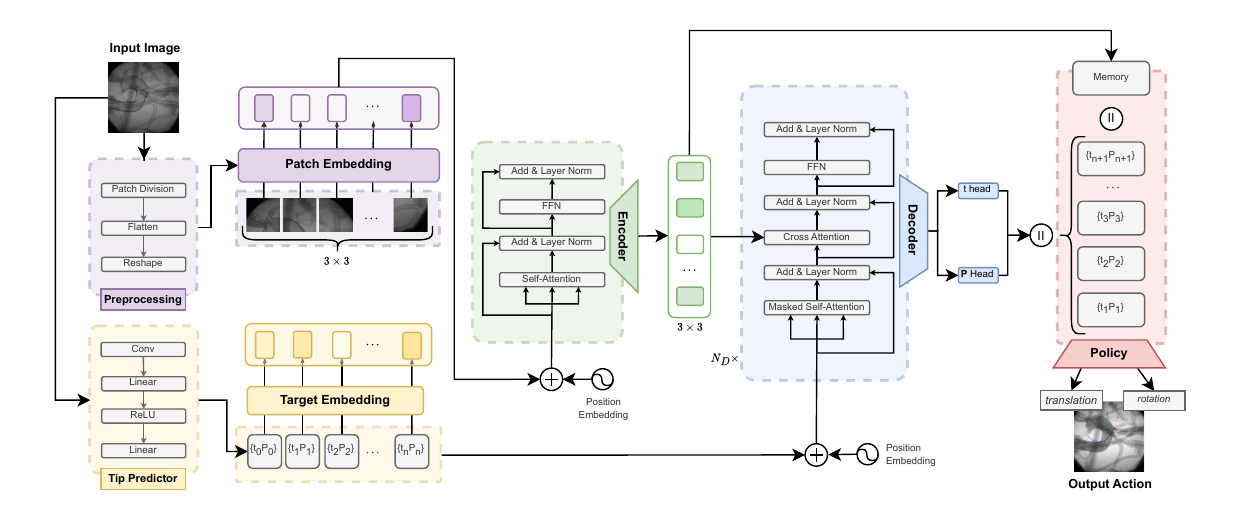}
    \vspace{0ex}
    \caption{\textbf{Network Architecture:} The input fluoroscopic image \( X \) is processed by a visual transformer encoder that divides the image into patches, embeds them, and generates visual feature representations \( X' \). Simultaneously, a positional encoder processes embeddings of the target sequence. These encoded features are fed into a transformer decoder composed of \( N_D \) layers. Using masked self-attention and cross-attention mechanisms, the decoder sequentially generates the B-spline control points \( \mathbf{P}_i \) and knots \( t_i \) that define the guidewire's geometry. The decoder predicts these parameters by projecting its outputs onto the B-spline dimensionality, yielding pairs \( \{\mathbf{P}_0, t_0\}, \{\mathbf{P}_1, t_1\}, \dots, \{\mathbf{P}_n, t_n\} \). An independent tip predictor module initializes the generation by predicting the starting point \( \{\mathbf{P}_0,t_0\} \).}
    \label{fig:main_architecture}
\end{figure*}

Accurate and efficient representation of curvilinear surgical tools (\ie guidewires and catheters) in interventional procedures is crucial for ensuring precise navigation and control. We choose a spline-based representation of the guidewire due to its explainability. This approach offers several distinct advantages due to its inherent smoothness and flexibility in modeling curvilinear structures~\cite{bismuth2012curvilinear}. The spline model, characterized by a set of control points and associated knots, allows for a compact, continuous, and smooth representation that accurately reflects the physical properties of the guidewire. Moreover, predicting the spline coefficients and knots directly enables an end-to-end approach that circumvents the need for intermediate segmentation steps, that is frequently adopted~\cite{spinelli2018first}. Our method not only enhances computational efficiency but also improves robustness to image noise and artifacts, while maintaining consistency across varying imaging conditions (\ie dimensionality-invariant representation). Lastly, the ability to predict spline parameters directly from image data facilitates real-time applications, as it reduces the processing time and potential errors associated with traditional segmentation-based approaches. Our network architecture can be visualized in Fig.~\ref{fig:main_architecture}.

\subsection{Spline Representation} 

Within fluoroscopic imaging, a guidewire can be ultimately represented as a B-spline~\cite{7381624}. A B-spline is a spline with minimal support \textit{w.r.t.} degree $p$ whose curve is composed of piecewise polynomial functions, parametrically defined over a set of control points and a non-decreasing sequence of knots. The curve \( \mathbf{C}(t) \) of degree \( p \) is given by:

\begin{equation}\label{eq:spline}
   \mathbf{C}(t) = \sum_{i=0}^{n} \mathbf{P}_i B_{i,p}(t)\,, 
\end{equation}
where \( \mathbf{P}_i \) represents the control points and \( B_{i,p}(t) \) are the B-spline basis functions of degree \( p \), defined on the knot vector \( t_0, t_1, \dots, t_m \). The basis functions \( B_{i,p}(t) \) are defined recursively using t recursion formula. Starting with the piecewise constant functions for \( p = 0 \):

\begin{equation}
  B_{i,0}(t) = 
\begin{cases} 
1 & \text{if } t_i \leq t < t_{i+1}, \\
0 & \text{otherwise},
\end{cases}  
\end{equation}
The higher-degree basis functions are recursively defined as:
\begin{equation}
    B_{i,p}(t) = \frac{t - t_i}{t_{i+p} - t_i} B_{i,p-1}(t) + \frac{t_{i+p+1} - t}{t_{i+p+1} - t_{i+1}} B_{i+1,p-1}(t).
\end{equation}

\begin{figure}
    \centering
     \includegraphics[width=\linewidth]{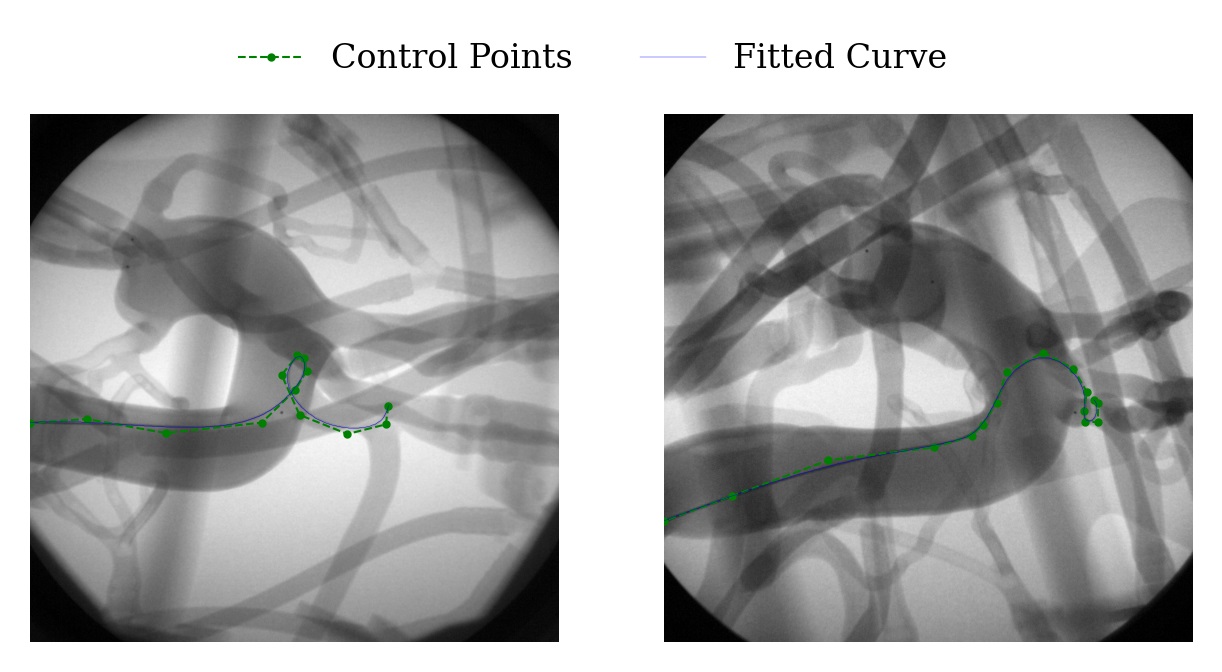}
     \vspace{-1ex}
     \caption{\textbf{B-spline Representation of a Guidewire:} The figure illustrates how a guidewire, as seen in fluoroscopic imaging, can be represented as a continuous B-spline curve.}
     \label{fig_spline}
\end{figure}

These functions \( B_{i,p}(t) \) are non-zero only within the interval \( [t_i, t_{i+p+1}) \), ensuring local control of the curve, which means that changes in a control point \( \mathbf{P}_i \) affect only the portion of the curve within this interval. Additionally, the constraint
$ \sum_{i=0}^{m-p-1} B_{i,p}(t) = 1\,$
for all \( t \) within the internal knots \( t_{p} \leq t \leq t_{m-p} \), normalizes the basis functions. To represent the guidewire, the parameter \( t \) represents the normalized arc length along the guidewire, providing a smooth and continuous model of the guidewire's geometry. Fig.~\ref{fig_spline} shows the B-spline guidewire representation.

\subsection{Network Architecture}
\textbf{Encoder.} Two encoders are used, with the first encoder resembling a visual transformer~\cite{dosovitskiy2020image}, where the image \(X \in \mathrm{R}^{H \times W }\) is divided into \(N_P\) patches, where \(N_P = \frac{H}{|P|}\times \frac{W}{|P|}\), and \(|P|\) represents the patch size. Subsequently, the patches are flattened into a 1D tensor. A linear embedding layer is then used to map the patches into a latent space. The second encoder is a sinusoidal positional encoder~\cite{vaswani2017attention} on the patches \(P\) and the target sequence $Y_{\text{TGT}} \in \mathbb{R}^{S \times E}$ with $S, E$ representing the length of the sequence and embedding dimensions, respectively.

\begin{table*}[t]
\begin{minipage}[b]{0.7\linewidth}
\centering
\resizebox{\linewidth}{!}{
\setlength{\tabcolsep}{0.2 em} 
{\renewcommand{\arraystretch}{1.8}
\begin{tabular}{c|c|c|cc|cc}
\hline
\multirow{2}{*}{\textbf{Method}} & \multirow{2}{*}{\textbf{Setup}} & \multirow{2}{*}{\textbf{Explainable?}} & \multicolumn{2}{c|}{\textbf{BCA}} & \multicolumn{2}{c}{\textbf{LCCA}} \\ \cline{4-7} 
 &  &  & \multicolumn{1}{c|}{\textit{Success (\%)}} & \textit{Time (\text{s})} & \multicolumn{1}{c|}{\textit{Success (\%)}} & \textit{Time (s)} \\ \hline
\begin{tabular}[c]{@{}c@{}}Expert  Teleoperation\end{tabular} & \begin{tabular}[c]{@{}c@{}}Fully Manual\end{tabular} & \_ & \multicolumn{1}{c|}{100} & \(32.1\pm9.1\) & \multicolumn{1}{c|}{100} & \(25.0 \pm 6.7\) \\ \hline \hline
\begin{tabular}[c]{@{}c@{}}GAIL-PPO\cite{chi2020collaborative}\end{tabular} & \begin{tabular}[c]{@{}c@{}}Semi-Autonomous\end{tabular} & No & \multicolumn{1}{c|}{69.4} & \(52.1\pm9.9\) & \multicolumn{1}{c|}{72.2} & \(76.5\pm24.1\) \\ \hline
\hline
\begin{tabular}[c]{@{}c@{}}Behavior Cloning~\cite{chi2020collaborative}\end{tabular} & \begin{tabular}[c]{@{}c@{}}Fully Autonomous\end{tabular} & No & \multicolumn{1}{c|}{5.60} & \(>200\) & \multicolumn{1}{c|}{\_} & - \\ \hline
\rowcolor[HTML]{EFEFEF}\begin{tabular}[c]{@{}c@{}}\textbf{SplineFormer} \textbf{(ours)}\end{tabular} & \begin{tabular}[c]{@{}c@{}}Fully Autonomous\end{tabular} & Yes & \multicolumn{1}{c|}{50.0} & \(150 \pm 45.6\) & \multicolumn{1}{c|}{\_} & - \\ \hline
\end{tabular}
}}
\vspace{5ex}
    \caption{Endovascular navigation results.
    }
    \label{tab:NavResult}
\end{minipage}
\hfill
\begin{minipage}[b]{0.3\linewidth}
\centering
\includegraphics[width=0.8\textwidth,height=0.9\linewidth]{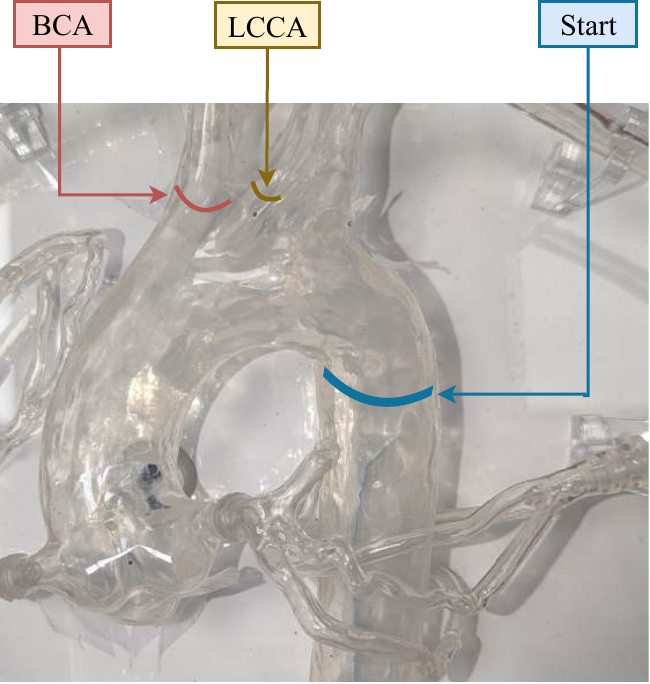}
\vspace{2 ex}
\captionof{figure}{Navigation setup.}
\label{fig:target-setup}

\end{minipage}
\end{table*}

The transformer encoder block consists of \(N_e\) stacked identical layers. Each of the layers adopted in the encoder consists of a multi-head self-attention (MHA) layer followed by a positional feed forward neural network. The MHA contains \(H\) parallel heads and each head \(h_i\) corresponds to an independent scaled dot-product attention function. This allows the model to jointly attend different sub-spaces. A linear transformation is the used \(W^O\) to aggregate the attention results of different heads. This is formulated as:
\begin{equation}
    \operatorname{MHA}(Q,K,V) = \operatorname{Concat}(h_1, h_2, \dots, h_H)W^O
\end{equation}

The scaled dot-product attention can then be computed as:

\begin{equation}
    \operatorname{Attention}(Q,K,V) = \operatorname{Softmax}\frac{QK^T}{\sqrt{d_k}}V
\end{equation}
where \(\{Q,K,V\}  \in \mathrm{R}^{N_q \times d_k}\), are the query, key and value matrix, respectively.

Following the self-attention layer, the positional feed-forward is implemented as two linear layers with GELU activation function, where a dropout is inserted between them. Mathematically, the layer is represented as:

\begin{equation}
    \operatorname{FFN}(x)=\operatorname{FC_2}((\operatorname{Dropout}(\operatorname{GELU}(\operatorname{FC_1}(x))))
\end{equation}

In each sublayer, there is a sublayer connection composed of a residual connection, followed by layer normalization:

\begin{equation}
    x^{out}= \operatorname{LayerNorm}(x^{in} + \operatorname{Sublayer}(x^{in}))
\end{equation}

\textbf{Decoder.} The decoder, composed of \(N_D\) transformer layers, sequentially generates the B-spline coefficients and corresponding knots that define the guidewire's geometry. Each layer includes masked self-attention, multi-head cross-attention, and a feed-forward network~\cite{vaswani2017attention}. The process begins with a target embedding layer, which maps input features to a hidden dimension, followed by sinusoidal positional encodings to capture the order of the sequence. The masked self-attention mechanism ensures autoregressive generation by restricting attention to previous or current coefficients, enforcing a sequential prediction. The final decoder output is projected onto the B-spline dimensionality, producing pairs \( \{P_0, t_0\}, \{P_1, t_1\}, \dots, \{P_n, t_n\} \), which define the curve \(\mathbf{C}(t)\) (Eq.~\ref{eq:spline}). The iterative generation of coefficients \(\mathbf{P}_i\) and knots \(t_i\), along with the end-of-sequence predictor \(s_i\), enables the model to capture the guidewire's complex curvilinear structure. A similar process has been used in text to speech works~\cite{chen2020multispeech}. This approach supports variable-length output sequences, facilitating the accurate modeling of intricate guidewire geometries.

\textbf{Tip Prediction.} Traditional transformer architectures rely on the start-of-sequence token~\cite{liu2021cptr} for initializing the generative process. Since the initial token represents the prediction anchoring point, we choose to dedicate an independent predictor composed of a combination of Convolutional Layers along with a ReLU activation and further linear layers to map the image to the target embedding. The predicted initial point is then used to generate the rest of the tokens. 

\subsection{Loss Function}
The global loss function $\mathcal{L}$ is composed of three loss functions, each assigned with a weighting $\lambda$ parameter: \textit{i)} A Mean Squared Error loss is used on the predicted sequence and the target sequence; \textit{ii)} a Binary Cross Entropy loss is used on the end-of-sequence prediction and its ground truth; and \textit{iii)} a curvature consistency loss by sampling the curve at \( n \) parameter values \( t_k \), where we select \( t_k \) uniformly distributed over the valid parameter range \( [t_p, t_{m-p}] \). \( t_p \) and \( t_{m-p} \) are the knots corresponding to the start and end of the curve segment influenced by the control points. The parameter values \( t_k \) are given by $t_k = t_p + \frac{k}{n-1}(t_{m-p} - t_p), \quad k = 0, 1, \dots, n-1 $. The total training loss is defined as following:




\begin{equation}\label{eq:loss_fn}
\begin{split} 
\mathcal{L} = \frac{1}{N} \sum_{i=1}^N \bigg(
& \lambda_a \big(\|t_i - \hat{t}_i\|^2 + \|\mathbf{P}_i - \hat{\mathbf{P}}_i\|^2\big) +  \\
& \lambda_b \big( -\mathbf{s}_i \log(\hat{\mathbf{s}}_i) - (1 - \mathbf{s}_i) \log(1 - \hat{\mathbf{s}}_i) \big) +\\
& \lambda_c \left\| \mathbf{C}(t_k) - \mathbf{\hat{C}}(t_k) \right\|^2 \bigg). 
\end{split}
\end{equation}

\section{Experiments} \label{Sec:exp}

We begin our experiments by training the shape estimation network on the dataset introduced in Section~\ref{sec:robotic-setup}. After training, we evaluate the learned policy on the state-action pairs using our endovascular robot. All experiments were conducted using a system consisting of an NVIDIA RTX 4080 GPU, \(\qty{128}{\giga\byte}\) RAM, and an Intel Core i9-13900 processor. The experiments were implemented using PyTorch.

\subsection{Autonomous Navigation Results}

We evaluated the performance of our proposed model, SplineFormer, in the task of autonomous guidewire navigation within a vascular environment. Starting from a predefined position within the descending aorta, we aimed to reach two distinct arterial targets: the brachiocephalic artery (BCA) and the left common carotid artery (LCCA), as depicted in Fig.~\ref{fig:target-setup}. For each target, we conducted \(20\) trials, recording the trajectories to create datasets of state-action pairs \((s_t, a_t)\). The agent received image-based X-ray observations, with the action space defined as \(a_t \in [-1,1]^2\), corresponding to a maximum translation of \(2\unit{mm}\) and a rotation of \(15\unit{\degree}\).

After training on the collected data, SplineFormer was deployed on our robotic platform. In a fully autonomous navigation task towards the BCA, it achieved a success rate of \(50\%\), with a mean completion time of \(2.5 \pm 0.76\unit{\min}\) over \(20\) trials. This performance marks a significant improvement over the baseline Behavior Cloning method~\cite{chi2020collaborative}, which attained only a \(5.6\%\) success rate in a fully autonomous setting. While our model did not outperform the semi-autonomous GAIL-PPO approach \cite{chi2020collaborative}, which had success rates of \(69.4\%\) for the BCA and \(72.2\%\) for the LCCA, it offers the advantages of full autonomy and explainability. Notably, fully autonomous endovascular navigation methods such as our SplineFormer and Behavior Cloning~\cite{chi2020collaborative} are unable to successfully cannulate the LCCA, highlighting areas for future enhancement.

\subsection{SplineFormer for Shape Prediction}\label{Subsec:splineformer} 
\begin{figure}[t]
     \centering
     \includegraphics[width=0.48\linewidth]{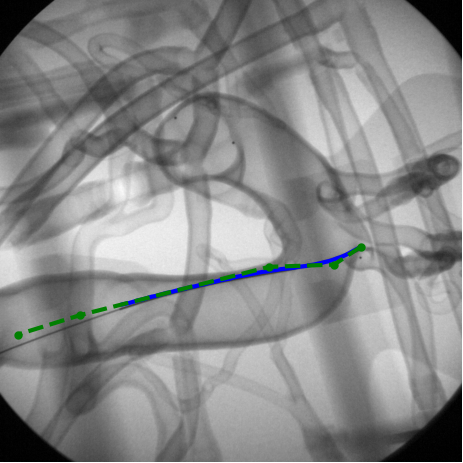}\hfill
   \includegraphics[width=0.48\linewidth]{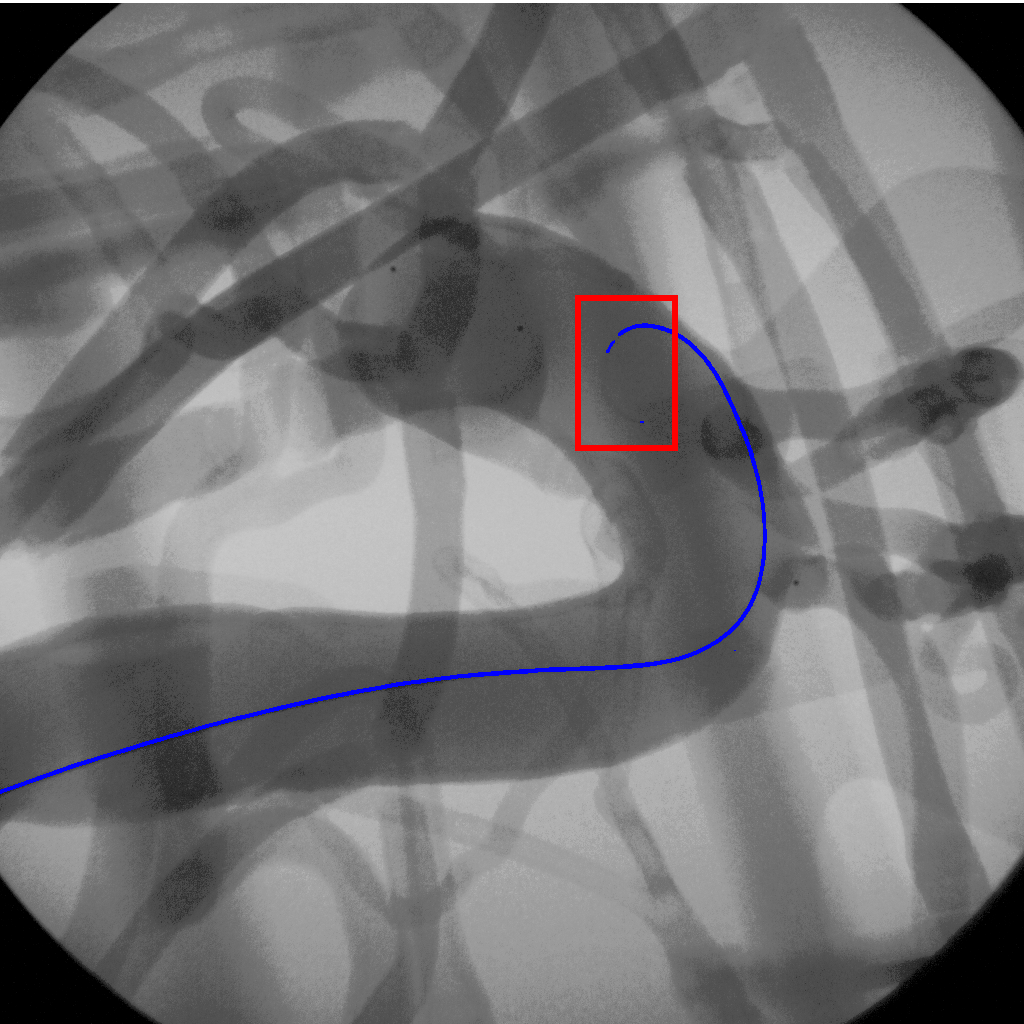}\\
     \vspace{1ex}
     \includegraphics[width=0.48\linewidth]{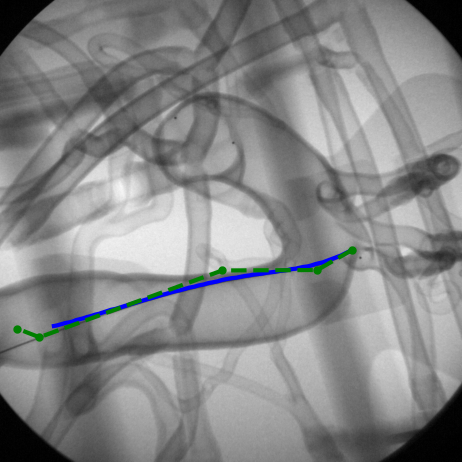}\hfill \hspace{1ex}
     \includegraphics[width=0.48\linewidth]{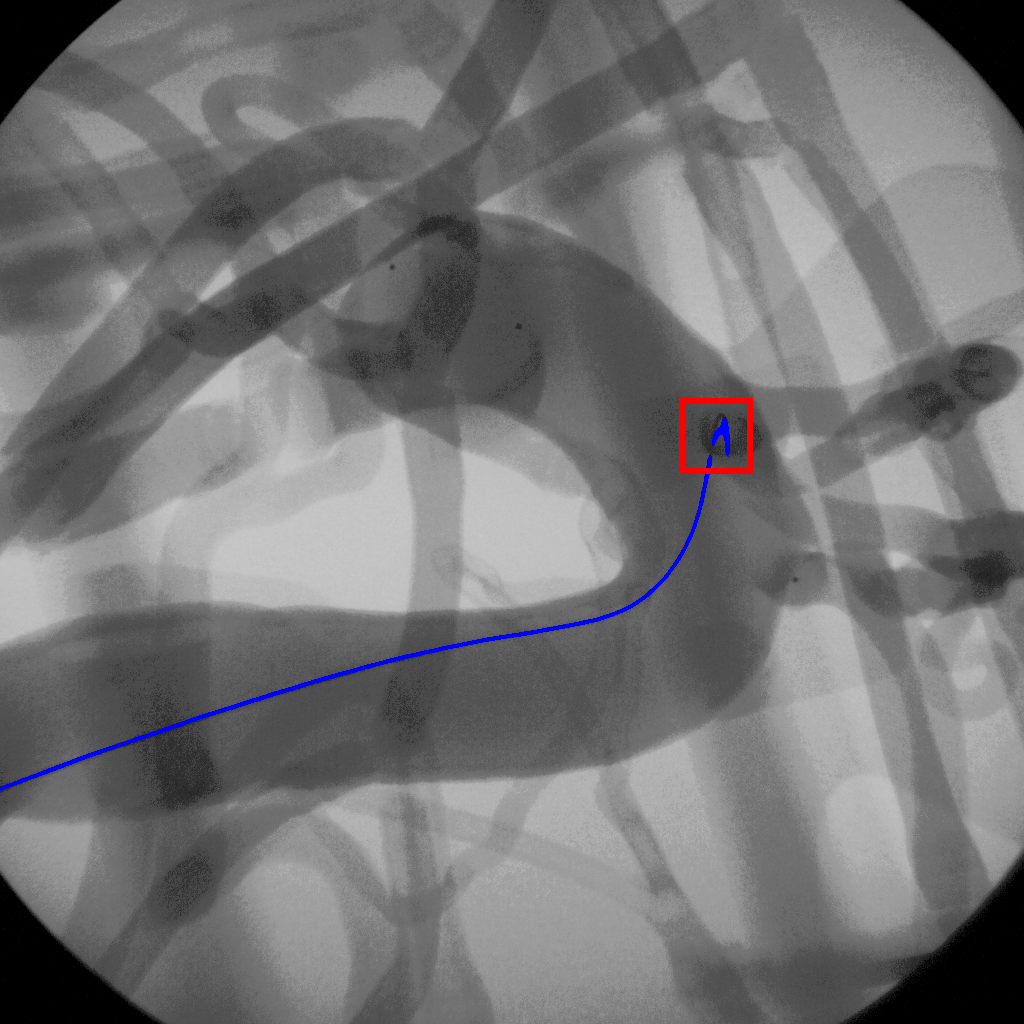}
     \vspace{1ex}
     \caption{\textbf{Comparison between SplineFormer and segmentation method:} SplineFormer (\textit{left column}) directly predicts the guidewire's geometry using a condensed information space. In contrast, U-Net generates segmentation masks without providing detailed geometric understanding, requiring additional processing for auxiliary tasks (\textit{right column}).}
     \label{fig:spline-vs-segment}
\end{figure}

We trained our network, SplineFormer, on the annotated dataset described in Section~\ref{sec:robotic-setup}. The model was trained for 300 epochs using the Adam optimizer with an initial learning rate of \(1 \times 10^{-5}\). The results, presented in Fig.~\ref{fig:spline-vs-segment}, illustrate that SplineFormer successfully predicts the overall shape of the guidewire within a compressed feature space. While SplineFormer is not explicitly designed for segmentation, it demonstrates a strong ability to detect key points of the guidewire, particularly near the tip. In contrast, traditional segmentation models such as U-Net~\cite{ronneberger2015u} generate qualitative masks but fail to capture the guidewire's detailed geometry, often missing the tip due to low contrast. Consequently, U-Net's segmentation outputs tend to be discontinuous and provide insufficient visual guidance for stable autonomous endovascular navigation.

\subsection{Attention Visualization} 
\begin{figure}[t]
    \centering
    \includegraphics[width=0.48\linewidth]{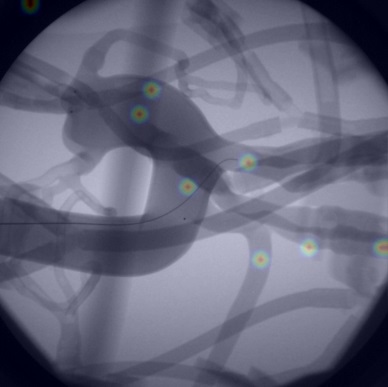}\hfill
    \includegraphics[width=0.48\linewidth]{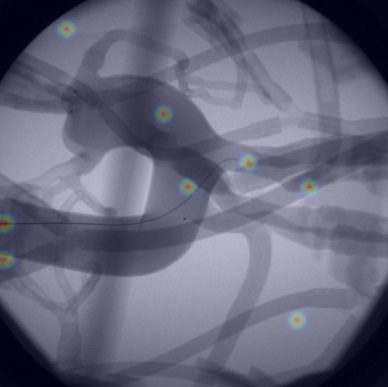}\\
    \vspace{1ex}
    \vspace{3ex}
    \caption{\textbf{Attention Visualization:} Attention maps generated through maximal fusion of attention heads~\cite{dosovitskiy2020image} highlight key regions where the guidewire tip is located, such as the central portion of the aortic arch, the ascending aorta, and the BCA cannulation target.}
    \label{fig:attention-maps}
\end{figure}

Building on the approach outlined in~\cite{dosovitskiy2020image}, we aggregate the attention heads using maximal fusion across the final layer, incorporating a discard factor to further isolate salient features. The resulting attention maps, shown in Fig.~\ref{fig:attention-maps}, highlight the model's focus on critical regions, specifically where the guidewire tip is located within the aortic arch. Unlike the attention maps generated by conventional segmentation methods, which often cover a big region, our approach concentrates attention on key points in the X-ray images, offering more targeted and precise localization.

\subsection{Discussion} \label{Subsec:disc}
We presented SplineFormer, a new approach to infer continuous surgical tool geometries by using our transformer-based model which utilizes B-splines to efficiently represent the geometries with minimal information. A strong focus of our exploration has been to reliably capture the entire geometry of the surgical tool (\ie guidewire) from the collated images. Indeed, one of the frequent shortfalls with traditional segmentation models like U-Net is the presence of discontinuity; a single row of pixels causes breakage and can render poorer performance. Intuitively, reconstructing a segmentation mask from the curve defined by the control points and knots of our B-spline leaves us with a deterministic, reproducible and complete curve. However, our work does not come without shortfalls. Given the parametrical representation, a small change in the curve parameters can yield a different curve and misrepresent the curve geometry. One such issue can be seen through an offset of the curve compared with the initial curve. Furthermore, while our trained SplineFormer model was able to successfully navigate and cannulate the BCA target on the physical robot, it struggled when attempting to reach the LCCA target. This was also observed in the previous method~\cite{chi2020collaborative} and is likely due to the more complex 3D anatomical structure of the aorta, which poses greater challenges for precise navigation. In practice, the BCA is typically the first major branch of the aortic arch and follows a more direct and accessible path, making it easier for robotic systems to navigate. In contrast, the LCCA is positioned as the second branch of the aortic arch, which curves and twists in a more intricate way, creating sharper turns and more challenging geometries for the model to handle in real robot experiments.




\section{Conclusions}\label{Sec:con}

We have proposed SplineFormer, a spline-based approach to autonomous endovascular navigation that condenses the guidewire's geometric information into a latent representation. Our method provides a more transparent and interpretable state-action representation, overcoming the limitations of traditional segmentation techniques constrained by the operating space. Our approach enhances flexibility and precision in navigation tasks. Future work will focus on refining our methodology by incorporating porcine-based environments and clinical data, aiming to bridge the gap between phantom experiments and clinical practice. We believe that leveraging explainable curvilinear representations has the potential to significantly advance autonomous endovascular navigation and ultimately improve patient outcomes.


\bibliographystyle{class/IEEEtran}
\bibliography{class/IEEEabrv,main}
   
\end{document}